\documentclass{article}
\usepackage{naturetex}

\usepackage{amsmath}
\usepackage{graphicx}
\usepackage{caption}
\usepackage{amssymb}
\usepackage{mathabx}
\usepackage{longtable}
\usepackage{lineno}

\begin{document}

\maketitle

{\setstretch{1.0}
	\section*{Abstract}
	\textbf{Stellar members of binary systems are formed from the same material, therefore they should be chemically identical. However, recent high-precision studies have unveiled chemical differences between the two members of binary pairs composed by Sun-like stars. The very existence of these chemically inhomogeneous binaries represents one of the most contradictory examples in stellar astrophysics and source of tension between theory and observations. It is still unclear whether the abundance variations are the result of chemical inhomogeneities in the protostellar gas clouds or instead if they are due to planet engulfment events occurred after the stellar formation. While the former scenario would undermine the belief that the chemical makeup of a star provides the fossil information of the environment where it formed, a key assumption made by several studies of our Galaxy, the second scenario would shed light on the possible evolutionary paths of planetary systems. Here, we perform a statistical study on 107 binary systems composed by Sun-like stars to provide - for the first time - unambiguous evidence in favour of the planet engulfment scenario. We also establish that planet engulfment events occur in stars similar to our own Sun with a probability ranging between 20 and 35$\%$. This implies that a significant fraction of planetary systems undergo very dynamical evolutionary paths that can critically modify their architectures, unlike our Solar System which has preserved its planets on nearly circular orbits. This study also opens to the possibility of using chemical abundances of stars to identify which ones are the most likely to host analogues of the calm Solar System.}

}

\section*{Introduction}\label{intro}

The observational evidence that planetary systems can be very different from each other suggests that their dynamical histories were very diverse \cite{Winn15}, probably as a result of a strong sensitivity to the initial conditions. Dynamical processes in the most chaotic systems have possibly destabilised planetary orbits, forcing them to plunge into the host star \cite{Mustill15}. Planet engulfment events involve the chemical assimilation of the planetary material into a star's external layer \cite{Pinsonneault01}. This causes a change in the chemical pattern of the stellar atmosphere in a way that mirrors the composition of the engulfed rocky object, with rocky-forming elements - such as iron - resulting more abundant than what they would be otherwise \cite{Cowley21,YanaGalarza16b,Chambers10}. 

Stellar associations, such as open clusters and binary systems, are the ideal targets to search for chemical signatures of planetary engulfment events. Their members have formed at the same time, within the same molecular cloud, and from the same material, therefore they are expected to be chemically identical. Unexpectedly, during the last few years an increasing number of high-precision spectroscopic studies have unveiled chemical differences among Sun-like stars belonging to the same association \cite{Ramirez15,Spina15,Teske16,Saffe17,Spina18b,TucciMaia19,Nagar20}. Although these chemical anomalies can be interpreted as the consequence of planetary engulfment events, they could also be explained as the result of a chemical segregation within the molecular cloud that gave origin to the stellar association \cite{Ramirez19}.

The solution to this ambiguity is expected to drive a generational advancement in astrophysics. In fact, unequivocal evidence of planet engulfment events and knowledge of their occurrence in Sun-like stars would shed light on the possible evolutionary paths of planetary systems, indicating how many of them have undergone complex phases of highly dynamical reconfiguration. Conversely, an evidence that molecular clouds are not chemically homogeneous would undermine the general belief that the chemical makeup of a star provides the fossil information of the environment where it formed, a key assumption made by several studies aiming to reconstruct the history of our Galaxy \cite{Freeman02}.

\section*{Results}\label{results}

In order to solve this puzzle, we perform a statistical study of a controlled sample of 107 binary systems composed by pairs of Sun-like stars with similar  effective temperatures T$_{\rm eff}$ and surface gravities log~g (a detailed description of our dataset is provided in Methods-A). We adopt atmospheric parameters and chemical abundances from multiple literature sources for 76 pairs \cite{Desidera04,Desidera06,Mack14,Liu14,Ramirez15,Teske16,Saffe17,Liu18,Oh18,Reggiani18,Saffe19,TucciMaia19,Nagar20,Hawkins20}. With the present work we enrich this existing dataset with the chemical analysis of stars in additional 31 pairs (methods and tools employed in our spectroscopic analysis are described in Methods-B), which are now part of our final sample of 107 binary systems.

\begin{figure}[h!]
\begin{center}
 \includegraphics[width=\textwidth]{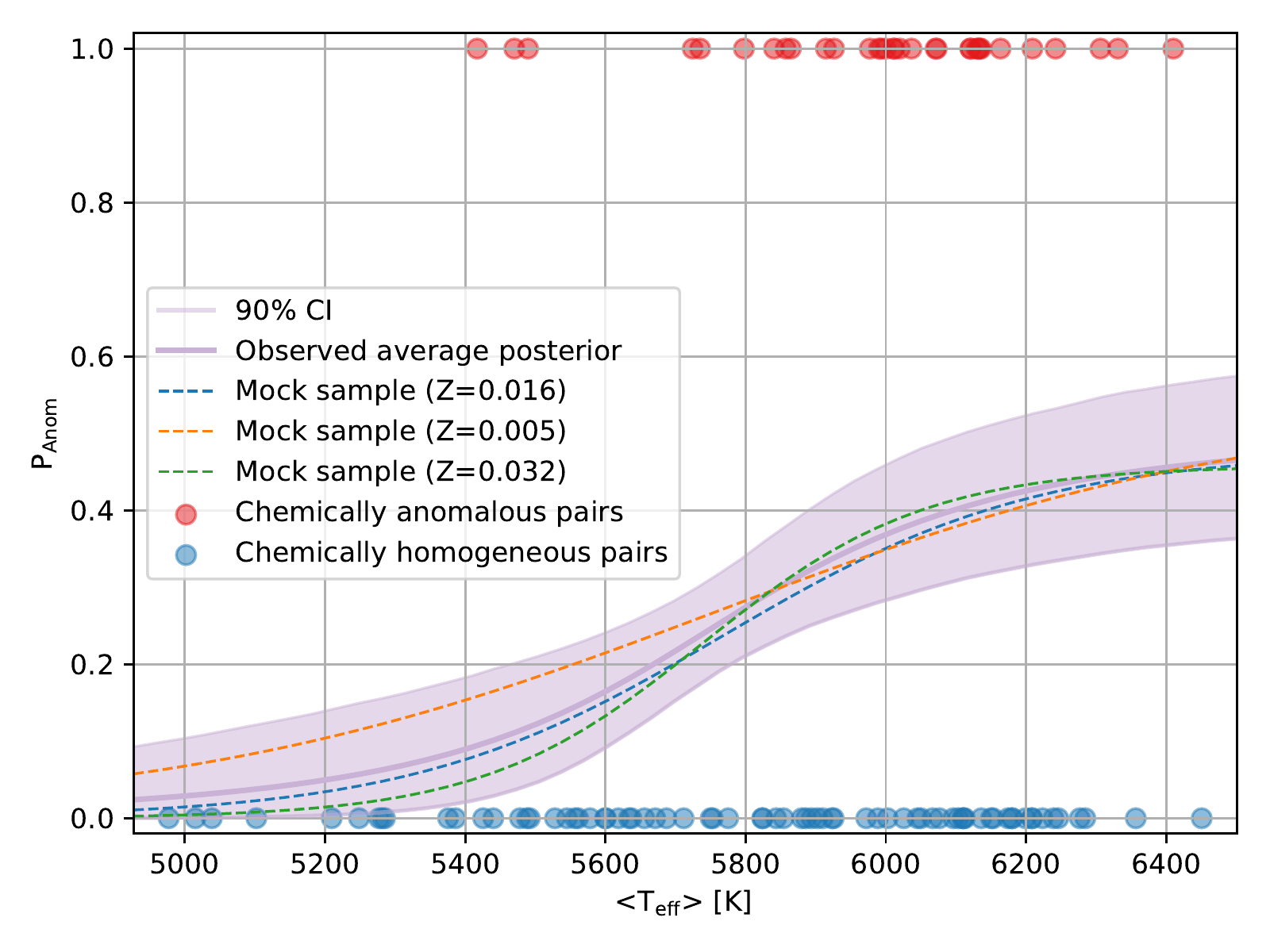}
 \end{center}
 \caption{\textbf{The frequency of \textit{chemically anomalous} pairs.} The binary systems are plotted as a function of the average temperature of their two components $\langle$T$_{\rm eff}$$\rangle$: the \textit{chemically homogeneous} pairs are plotted as blue circles along the lower x-axis, while the \textit{chemically anomalous} binaries as red circles along the upper x-axis. The probability of finding \textit{chemically anomalous} pairs P$_{\rm Anom}$ is modelled on these observations: the resulting 90$\%$ confidence interval and average of the posteriors are shown as a purple shaded area and a purple solid line, respectively. The three dashed lines (blue, orange, and green) represent the probability estimates for the mock dataset calculated at different metallicities (Z = 0.016, 0.005, and 0.032). The dependence of chemical anomaly probability on stellar effective temperature supports the planet engulfment hypothesis.}
 \label{prob_estimate}
\end{figure}

Even though stars in binary systems are expected to share an identical chemical pattern, the stellar components of 33 pairs in our sample have iron abundances that are anomalously different at the two-sigma level. These binaries are hereafter labelled as \textit{chemically anomalous}. Conversely, the remaining 74 pairs are labelled as \textit{chemically homogeneous}. These two classes of stellar pairs are plotted in Figure~\ref{prob_estimate} as a function of the mean effective temperature of the two binary components $\langle$T$_{\rm eff}$$\rangle$. More specifically, the \textit{chemically homogeneous} pairs are plotted as blue circles along the lower x-axis, while the \textit{chemically anomalous} binaries are the red circles along the upper x-axis. Interestingly, a Markov chain Monte Carlo simulation conditioned on this dataset indicates that the probability of finding a \textit{chemically anomalous} binary P$_{\rm Anom}$ is a function of $\langle$T$_{\rm eff}$$\rangle$ (for more details on the model and the statistical analysis, see Methods-C). This is readily evident from Figure~\ref{prob_estimate} which shows the P$_{\rm Anom}$ posterior average (purple solid line) and its 90$\%$ confidence interval (purple shadowed area) as a function of $\langle$T$_{\rm eff}$$\rangle$: while the P$_{\rm Anom}$ values are nearly zero at low $\langle$T$_{\rm eff}$$\rangle$, they increase towards the hottest pairs.  

Although this is the first time that P$_{\rm Anom}$ is examined against the typical effective temperature of the binary components and that a positive correlation between the two variables is observed, this relation is an expected outcome of planet engulfment events \cite{Pinsonneault01}. In fact, Sun-like stars with lower T$_{\rm eff}$ have thicker convective zones that can easily dilute a large amount of rocks acquired from their planetary systems without changing their chemical composition. However, the mass enclosed in the external layer of Sun-like stars shrinks as we move towards the hottest atmospheres, till the point that these latter have an extremely thin convective zone that can easily get contaminated by external material. Therefore, the hottest Sun-like stars of our sample will record almost any planet falling into their atmospheres as a variation of their chemical composition. As a result, the hotter the star, the higher the probability of observing chemical anomalies due to planet engulfment events.

According to our model, the mean probability of finding a \textit{chemically anomalous} pair for the hottest binaries (i.e., T$_{\rm eff}$~=~6500~K) is 0.47, with a 90$\%$ confidence interval of 0.36-0.58. Assuming that these atmospheres are so thin that any amount of planetary material would change their composition, and assuming that a pair becomes \textit{anomalous} when at least one of the two components has swallowed a planet, we can define the probability of planet engulfment events in Sun-like stars as P$_{\rm Engulf}$=1$-$$\sqrt{1-P_{\rm Anom}}$. Therefore, based on the results of our model the mean P$_{\rm Engulf}$ is 0.27 with a 90$\%$ confidence interval equal to 0.20-0.35.

As an exemplification of what we have expressed above, we produce a mock sample of binary pairs whose components have the same probability of ingesting planetary material that we derived from the observed dataset (i.e., P$_{\rm Engulf}$ = 0.27). The amount of material that is swallowed by each star is randomly drawn from a Gaussian distribution centred at  2.0 M$_{\Earth}$, with a standard deviation of 1.0 M$_{\Earth}$ and ranging between 0 and 10 M$_{\Earth}$. We repeat the analysis over three sets of stellar models \cite{Spada17} characterised by different metallicities: solar metallicity (Z$_{\odot}$=0.016), metal-rich (Z=0.032), and metal-poor (Z=0.005). We also assume that the planetary material diluted into the stellar convective zone has the same abundance distribution of the metals as that observed in the Earth and a global metallicity that scales with the metallicity of the hosting star (Z$_{\rm planet}$ = Z$_{\Earth}$$\times$Z/Z$_{\odot}$). The average probabilities estimated from these mock samples are shown in Figure~\ref{prob_estimate} as coloured dashed lines and they closely follow the probabilities inferred from the observed dataset. This demonstrates that the observed distribution of \textit{chemically homogeneous} and \textit{anomalous} pairs as a function of their $\langle$T$_{\rm eff}$$\rangle$ is what we expect being the result of planet engulfment events.

\begin{figure}[h!]
\begin{center}
 \includegraphics[width=\textwidth]{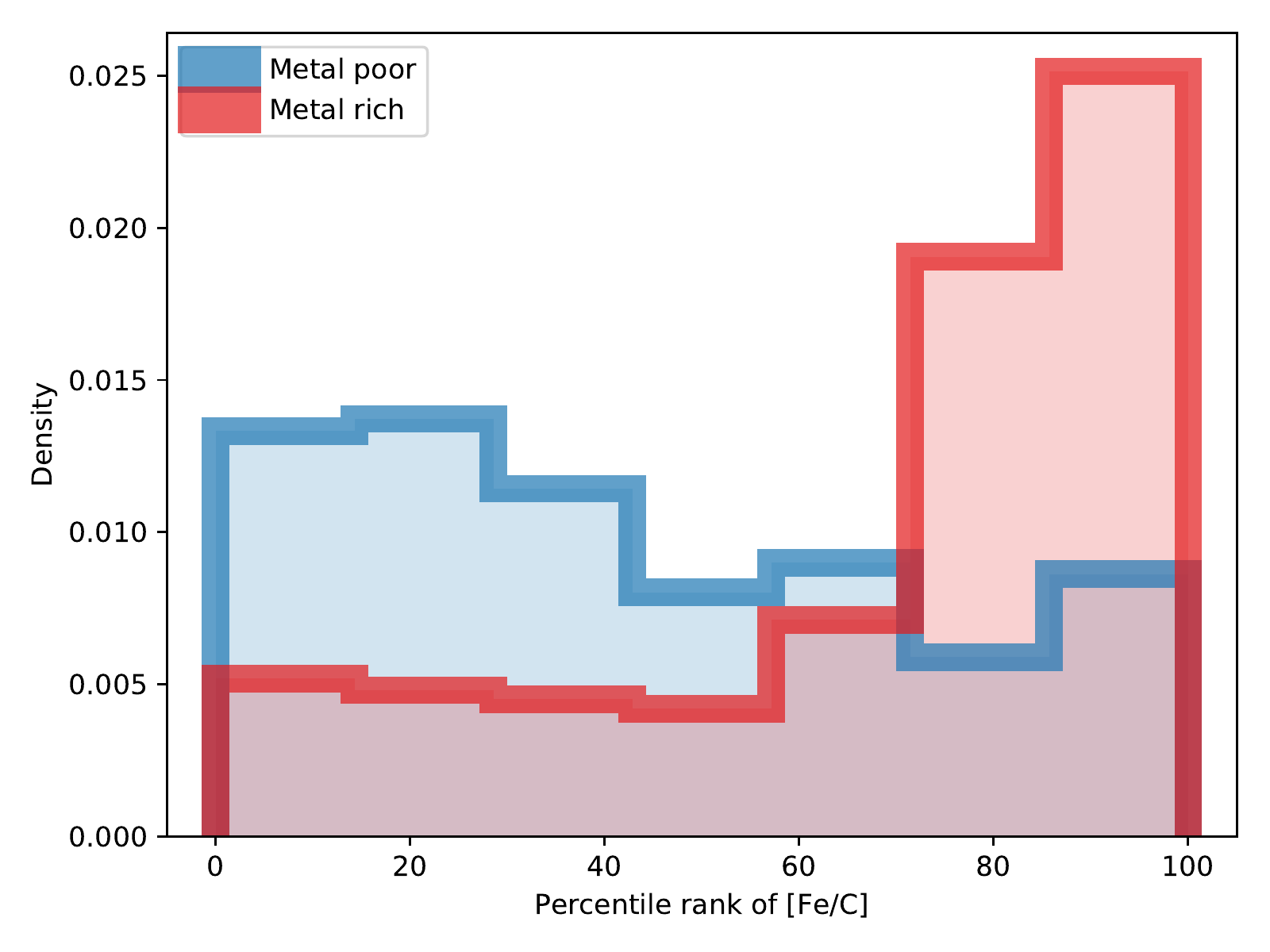}
  \end{center}
 \caption{\textbf{The [Fe/C] percentile ranks.} The plot shows the distribution of [Fe/C] percentile ranks for the metal-rich (red histogram) and metal-poor (blue histogram) components of \textit{chemically anomalous} pairs. The high [Fe/C] percentile ranks characteristic of the metal-rich components indicate that these stars have [Fe/C] ratios that are typically larger than the values seen in solar twin stars.}
 \label{volatiles_refractories}
\end{figure}

When planetary material enters the star and pollutes its convective zone, the stellar atmospheric composition changes in a way that mirrors the composition observed in rocky objects, namely, with refractory elements being more abundant than volatiles. Therefore, stars that have engulfed planetary material should have higher abundance ratios of refractories over volatiles than the typical values found in stars of similar ages and metallicities. In Figure~\ref{volatiles_refractories} we test this hypothesis using iron as a proxy of the refractory elements and carbon for the volatiles. The plot shows the percentile ranks of [Fe/C] relative to a control sample of Sun-like stars for the the metal-rich (in red) and metal-poor (in blue) components of the \textit{chemically anomalous} pairs (for more details on the analysis, see Methods-D). When a star has a large percentile rank, this indicates that its [Fe/C] ratio is higher than the typical values found in the control sample. Conversely, a star with a small percentile rank has a small [Fe/C] ratio compared to the control sample. As we expect from the planet engulfment scenario, the distribution of the metal-rich components peaks at high percentile rank values, indicating that they are typically richer in refractories (i.e., iron) relative to the control sample, but not in volatiles (i.e., carbon). On the other hand, the metal-poor components are more uniformly distributed along most of percentiles range, meaning that their abundance ratios of refractories over volatiles are similar to the typical values seen in nearby solar twin stars. We also observe that the distribution of the metal-poor components slightly decreases towards large percentile rank. This behaviour may reflect the fact that the metal-poor components have retained their original proportion of refractory elements relative to volatiles, while a fraction of the control sample is expected to have undergone planet engulfment events resulting anomalously enriched in refractories.

\begin{figure}[h]
\begin{center}
 \includegraphics[width=\textwidth]{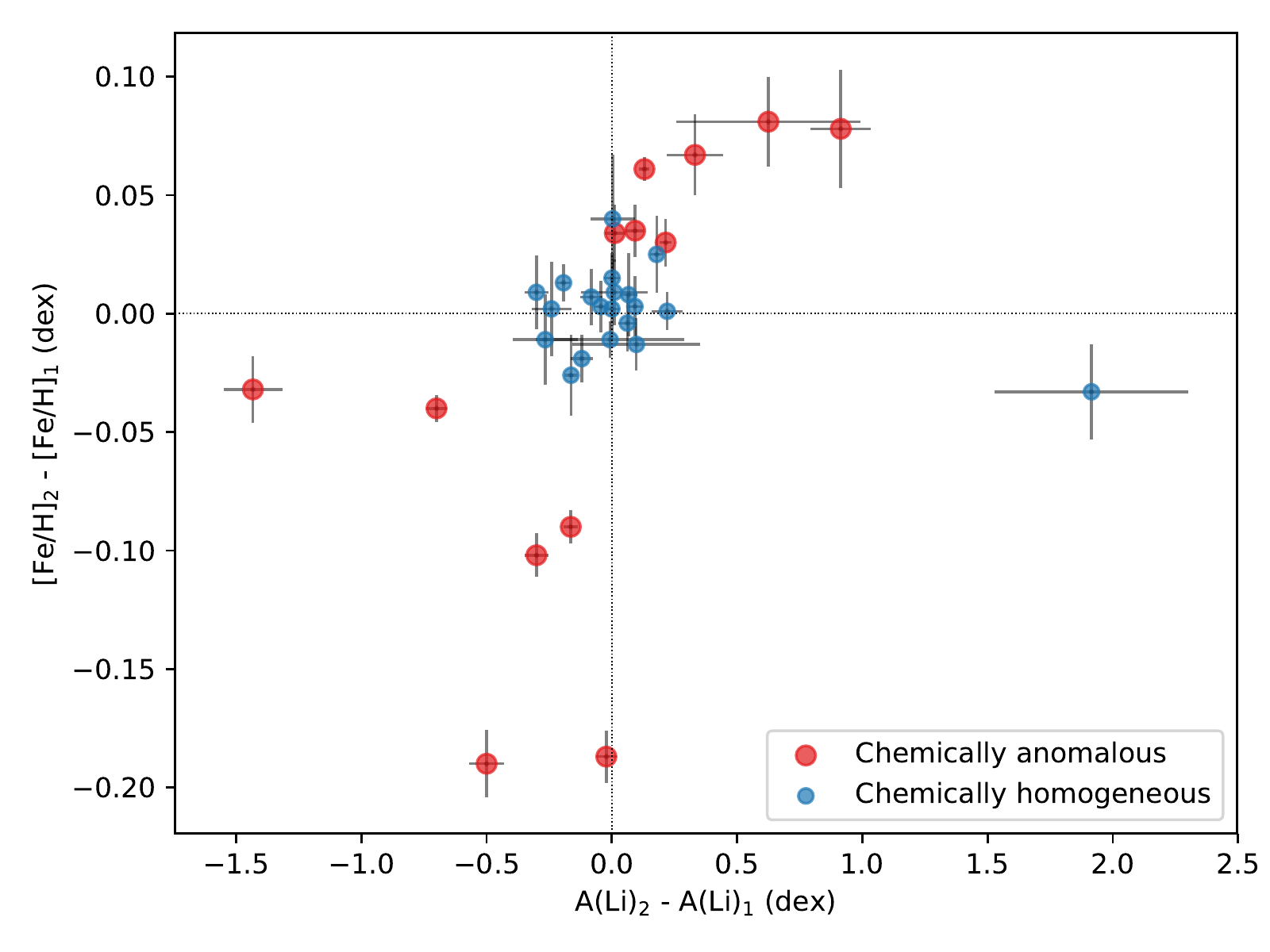}
  \end{center}
 \caption{\textbf{The $\Delta$Fe-$\Delta$Li diagram.} The differential iron abundances $\Delta$Fe of the binary pairs are plotted as a function of their differential lithium abundances $\Delta$Li. The \textit{chemically homogeneous} pairs are shown in blue, while the \textit{chemically anomalous} are in red.}
 \label{Lithium}
\end{figure}

Under certain conditions, planet engulfment events can also lead to a significant increase in the stellar lithium abundance \cite{Sandquist02}. All stars are born with a similar amount of Li in their atmospheres. However, Sun-like stars burn most of this initial endowment during theirs first few 100 Myr. Afterwards, this Li depletion will significantly slow down, even though it is expected to continue at a very small regime during the entire main-sequence phase \cite{Carlos19}. If the engulfment event occurs when the star is old enough that it has already burnt most of its Li, the stellar atmosphere receives from the planet a new substantial supply of this element. This will lead to a considerable increase of the stellar Li abundance, which is expected to persist for a long time because the burning process is acting at a reduced speed. Conversely, if the star swallows the planet when it is too young and still contains most of its initial Li reservoir, the new Li supply carried by the planet is similar to a drop in the ocean that can hardly produce any significant variation of the Li abundance in the stellar atmosphere. The influence of planet engulfment events on Li abundances is clearly illustrated in Figure~\ref{Lithium}, which shows the differential iron abundance $\Delta$Fe between the two components of binary pairs as a function of their differential lithium abundance $\Delta$Li. We find that the \textit{chemically homogeneous} pairs (blue circles) are restricted within $\Delta$Li values of $-$0.4 and $+$0.3, with only one notable exception at $\Delta$Li=1.9$\pm$0.4 dex. On the other hand, the \textit{chemically anomalous} pairs (red circles) are spread over a much larger $\Delta$Li range. Indeed, most of the stars that are anomalously richer in iron relative to the companion are also richer in Li, which is easily explained by planet engulfment events. At the same time, there are no iron-rich components that are poorer in Li, which would represent a clear contradiction to the planet engulfment scenario. Instead, there are \textit{chemically anomalous} pairs that are homogeneous in Li. These latter can be explained by planet engulfment events occurred when the star was younger than a few 100 Myr. 

It must be noted that a planet engulfment event is not the only process that can imprint a significant difference of Li abundance between two stars in a binary systems. For example, stars with different masses can deplete Li at very different rates and this factor must be considered when using Li as chemical indicator of planet engulfment events. This is especially true for stars with $T_{\rm eff}$$<$6000 K \cite{Bensby18}. Therefore, we excluded from Figure~\ref{Lithium} all the pairs with $\langle$T$_{\rm eff}$$\rangle$$<$6000~K and composed by stars whose T$_{\rm eff}$ and log~g differ by more than 100 K and 0.1 dex, respectively. The remaining 42 binary systems with Li abundance determinations are those shown in Figure~\ref{Lithium}. These are the pairs that should naturally be homogeneous in Li abundance.

\newpage
\section*{Discussion}
In Figure~\ref{prob_estimate} we demonstrate that the probability of finding a \textit{chemically anomalous} binary increases with the average temperature of the pair. This result cannot be explained by hypothetical inhomogeneities of the protostellar cloud. Instead, it is evidence that the stellar convective zones have been polluted by external material, which has altered the atmospheric chemical compositions.

What kind of material is responsible of these pollution episodes? A clear evidence that this material consists in falling planets or planetesimals would allow us to relate the frequency of chemically anomalous stars to the demographics of exoplanetary systems and the dynamical processes that shape their architectures. However, these abundance variations could in principle be originated by other mechanisms, such as a selective accretion of gas or dust from the protoplanetary disks that stars harbour during their very early stages of life \cite{Onehag11,Gaidos15}. Therefore, how can we be sure about the nature of the polluting material?

Both classical and recent models of stellar evolution indicate that the convective zones of Sun-like stars are $\geq$50 per cent by mass \cite{Kunitomo18} during the time while they were surrounded by protoplanetary disks (i.e., $\leq$5-10 Myr; \cite{Mamajek09}). Such overly thick convective zones can dilute more than the entire iron contained in our Solar System without producing any significant variation of the atmospheric chemical patterns \cite{Spina15}. Therefore, the observed variation cannot be protoplanetary disk phenomenon. In this regard, we also demonstrate that the observations are consistent with a scenario where stars are being polluted by Earth-like material accreted from their planetary systems (see dashed lines in Figure~\ref{prob_estimate}).

The results outlined in Figure~\ref{volatiles_refractories} provide additional details on the nature of the polluting material, revealing that it is rich in refractory elements and poor in volatiles. Furthermore, although previous works have shown that the chemical anomalies in binary systems are generally characterised by different proportions in refractory elements between the two components of the pair \cite{Biazzo15,Nagar20}, they could not establish which one of the two stars is the one with an \textit{anomalous} composition. For the first time, we demonstrate that the more metal-rich components have systematically higher abundances of refractories than what is expected from coeval stars of similar metallicities, while the metal-poorer components have ordinary abundance patterns. This clue strongly indicts rocky material for being responsible of the pollution of the metal-rich components of the \textit{anomalous} pairs. The scenario is further confirmed by Figure~\ref{Lithium}, which indicates that the accreted material is - in most of the cases - significantly richer in Li than the unpolluted stellar atmosphere.

The pieces of evidence described above provide unambiguous demonstration that planet engulfment events occur in Sun-like stars, and that these episodes are able to alter the stellar chemical composition. Finding the cause of chemically anomalous Sun-like stars in binary systems resolves one of the most significant contradictions in modern stellar astrophysics and arises important implications for this field. For instance, if elemental abundances in 27$\%$ of Sun-like stars can be altered by planetary engulfment events, then stellar chemical patterns are no longer entirely reflective of the star's progenitor cloud. Therefore, these results have a direct impact on our ability of tagging stars to their birth environment simply based on their chemical composition, a goal that present and future large spectroscopic surveys aim to achieve. However, as we show in Figure~\ref{prob_estimate}, the chemical consequences of planetary ingestion are extremely sensitive to the thickness of the stellar external layer. Therefore, low-mass main sequence stars and giants are not expected to vary their chemical composition due to their extremely thick external layers. 

Our results also have significant implications related to exoplanet science. One of the most interesting discoveries over the short history of exoplanet exploration is that, although planetary systems are common in the Galaxy, they are in many ways quite different from each other \cite{Winn15}. This diversity is likely the result of an extreme sensitivity to the initial conditions that can lead planetary systems to extremely different evolutionary paths. All this is driven by severe dynamical processes that can impose significant reconfiguration of planetary systems architectures. Our study provides additional evidence that a non-negligible fraction of planetary systems around Sun-like stars have experienced an extremely dynamical past, culminating with the fall of planetary material into the host star. The critical aspect of novelty of our research is that it solely relies on a comprehensive description of stellar chemical patterns, thus it is completely independent from both exoplanet detection techniques, which can be heavily biased towards specific types of planets, and n-body numerical simulations, which are often anchored to the observed demography of planetary systems.

Besides the precious insights on exoplanets populations and their formation mechanisms, this study can also be of practical utility for the achievement of one of the greatest scientific challenges of our decade: the search of Earth twins. While astronomers have discovered over 4,000 exoplanets, none of these are like our own planet. This is because the detection of Earth-like planets orbiting Sun-like stars requires to surpass the precision barrier currently set by stellar variability at 10 cm/s. Furthermore, the timescale of the observations needed to find an Earth analog is multiple years at minimum. Unfortunately, even when these limitations are overcame, there are millions of nearby Sun-like stars that can be potentially observed. All that will make the search for Earth-like planets like looking for the proverbial ``needle in a haystack". The possibility to detect chemical signatures of planet engulfment events implies that we can use the chemical composition of a star to infer if its planetary system has undergone an extremely dynamical past, unlike our Solar System which has preserved its planets on nearly circular orbits with very limited migrations. Therefore, we now have a potential ``upstream'' method to identify those Sun-like stars that are more likely to host Earth-like planets. Would this method work for the Sun, the only Sun-like star that we know is hosting a Earth-like planet? The answer is promisingly positive. In fact the Sun was found to have an unusual and still unexplained chemical composition when compared to other Sun-like stars, with a lower content of refractory elements and lithium \cite{Melendez09,Bedell18,Carlos19}. This peculiar chemical composition of the Sun cold be linked to the distinctively ordered architecture of our Solar System.

\newpage
\section*{Methods}

{\centering
\subsection*{A. Dataset}
\label{Met:Dataset}
}

The dataset employed in this study is formed by 107 binary pairs of Sun-like stars. More specifically, we include pairs with $\langle$T$_{\rm eff}$$\rangle$~$<$ 6500 K, because warmer stars have thick radiative envelopes \cite{Pinsonneault01}, thus they are not suitable for our study. We also impose that the difference between the two stellar T$_{\rm eff}$ and log~g within each pair are smaller than 600 K and 0.6 dex, respectively. This guarantees that each pair is formed by stars with similar atmospheres, which reduces the impact of unknown systematics in the determination of $\Delta$Fe values and other abundances. We also verify that our scientific results hold also if we decrease the thresholds on the differential T$_{\rm eff}$ and log~g to 400 K and 0.4 dex, respectively.

Atmospheric parameters and abundances of stars in 31 binary systems are derived in this work from HARPS spectra. Most of these pairs are well know bound associations, all of them are systems confirmed through Gaia eDR3 astrometry. The spectroscopic analysis of these stars is described in the following subsection Methods-B. This dataset is then enriched with other 76 pairs taken from the literature. Among all the available spectroscopic studies of binary pairs formed by Sun-like stars, we only consider those that reached precisions in atmospheric parameters and chemical abundances that are typically comparable to those resulting from our analysis of HARPS spectra. 

We divide these works from the literature in two groups based on the quality and precision of their results:

\begin{itemize}
\item The first group is the one with the higher quality. This includes 12 pairs analysed by Nagar et al. (2020) \cite{Nagar20}, that are not in common with our HARPS sample. All these spectra were observed at resolving power R$\geq$80k and signal-to-noise ratios SNRs$\geq$300 pxl$^{-1}$. The method of analysis was based on a line-by-line differential analysis, which is the same technique adopted in the analysis of the HARPS sample. We also include 10 additional pairs analysed by other studies \cite{Mack14,Liu14,Ramirez15,Teske16,Saffe17,Liu18,Oh18,Reggiani18,Saffe19,TucciMaia19} relying on spectra of comparable quality and similar methods of analysis. The dataset this first group provides the iron abundances used in Figure \ref{prob_estimate} and - when available - also the abundances of Li and C employed in Figures \ref{volatiles_refractories} and \ref{Lithium}.

\item Second group. We added to our sample 20 of the binary systems from Hawkins et al. (2020) \cite{Hawkins20}, excluding those that do not satisfy our selection criteria and those that are already included in our HARPS sample or in the papers mentioned above. Similarly, we added 34 of the 56 binary systems analysed by Desidera et al. (2004, 2006) \cite{Desidera04,Desidera06}, excluding also those for which atmospheric parameters were determined with less than 100 Fe lines. From the second group we only use the iron abundances and we disregard abundance determinations of other elements because of their lower precision.
\end{itemize}

In Table~\ref{lit_data} we list the pairs included in our final dataset. For each of the 107 pairs we report the name of the two stellar components, their $\langle$T$_{\rm eff}$$\rangle$, $\Delta$Fe and the reference to the work that analysed the stars. Our and some other studies in the literature have calculated $\Delta$Fe and relative uncertainty through a differential line-by-line analysis of the iron lines of one star relative to the companion. However, when this information is not provided, we derive $\Delta$Fe for the two components A and B of a binary system as [Fe/H]$_{\rm A}$-[Fe/H]$_{\rm B}$. Furthermore we define uncertainty associated to $\Delta$Fe as $\sigma_{\Delta Fe}$ = $\sqrt{\sigma_{[Fe/H]_A}^2 + \sigma_{[Fe/H]_B}^2}$, where $\sigma_{[Fe/H]}$ is the uncertainty associated to the [Fe/H] abundance of a single star. 

\begin{figure}[h]
\begin{center}
 \includegraphics[width=\textwidth]{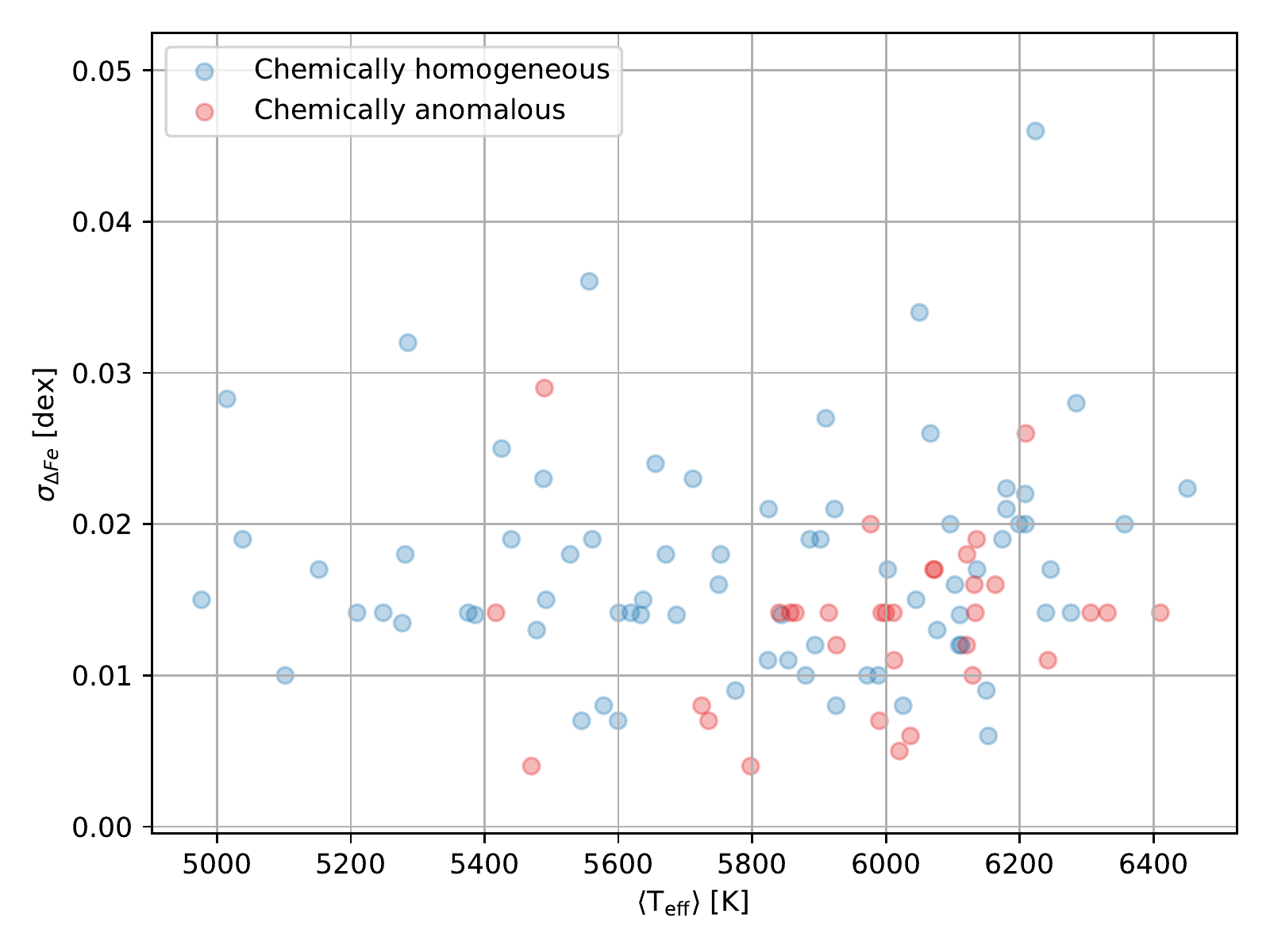}
  \end{center}
 \caption{\textbf{The $\sigma_{\rm \Delta Fe}$-$\langle$T$_{\rm eff}$$\rangle$ diagram.} The figure shows the uncertainty associated to the differential iron abundance $\sigma_{\rm \Delta Fe}$ as a function of the average temperature of the two binary components $\langle$T$_{\rm eff}$$\rangle$. The \textit{chemically homogeneous} pairs are shown in blue, while the \textit{chemically anomalous} are in red.}
 \label{checks}
\end{figure}

In our analysis we define as \textit{chemically anomalous} all the pairs that satisfy the following criterium:
\begin{equation}
\label{sigmoid}
   |\Delta Fe| > 2 \times max(\sigma_{\Delta Fe}, 0.01).
\end{equation}
Because of this definition, it is of paramount importance to verify the absence of any correlation between $\sigma_{\Delta Fe}$ and $\langle$T$_{\rm eff}$$\rangle$ that could introduce spurious variations of P$_{\rm Anom}$ as a function of $\langle$T$_{\rm eff}$$\rangle$. Figure~\ref{checks} shows that the $\sigma_{\Delta Fe}$ values are uniformly distributed across the entire range spanned by $\langle$T$_{\rm eff}$$\rangle$. As a further check, we computed a two-sample Kolmogorov-Smirnov test that evaluates whether the $\sigma_{\Delta Fe}$ values of pairs with $\langle$T$_{\rm eff}$$\rangle$$<$5600K and those of pairs with $\langle$T$_{\rm eff}$$\rangle$$>$5700K are drawn from the same distribution. The resulting p-value is equal to 0.84, indicating that there are no differences between the two groups.

{\centering
\subsection*{B. Spectroscopic analysis}
}

The 62 stars analysed in this work are observed with the HARPS spectrograph \cite{Mayor03} on the 3.6 m telescope at the La Silla observatory, mostly through our ESO Programmes (IDs 0103.C-0785 and 0101.C-0275). The spectra were acquired with a resolution of 110k, while the SNRs of the coadded spectra range between 100 and 900 pxl$^{-1}$, with a median value of 300 pxl$^{-1}$. In addition to the stars in binary systems, the sample includes a number of solar spectra of SNR of 1200 pxl$^{-1}$ acquired through HARPS observations (ESO Programme 188.C-0265) of the asteroid Vesta. 

The methods and tools used in our spectroscopic analysis are already tested in our previous works \cite{Nagar20,Casali20}. In short:
\begin{itemize}
\item Equivalent widths (EWs) of the atomic transitions listed in Mel\'{e}ndez et al. (2014) \cite{Melendez14} are measured with Stellar diff, a Python code publicly available at \url{https://github.com/andycasey/stellardiff}. This code is particularly indicated for differential analysis because it allows the user to hold the same assumptions in the choice of the local continuum around the lines of interest. This is expected to minimise the effects of an imperfect spectral normalisation or unresolved features in the continuum that can lead to larger errors in the differential abundances \cite{Bedell14}.

\item The iron EWs are processed by the qoyllur-quipu (q2) code \cite{Ramirez14} which performs a line-by-line differential analysis relative to the Solar spectrum and automatically estimates the stellar parameters (T$_{\rm eff}$, log~g, [Fe/H], and microturbulence v$_{\rm t}$) by iteratively searching for the excitation/ionisation equilibria of iron lines. We assumed the nominal solar parameters, T$_{\rm eff}$=5777 K, log~g=4.44 dex, [Fe/H]=0.00 dex and v$_{\rm t}$ =1.00 km s$^{-1}$ \cite{Cox00}. The iterations are executed employing the Kurucz (ATLAS9) grid of model atmospheres \cite{Castelli04}. In each step the abundances are estimated using MOOG (version 2014, \cite{Sneden73}). The errors associated with the stellar parameters are then evaluated by the code, which also takes into account the dependence between the parameters in the fulfilment of the equilibrium conditions \cite{Epstein10}. We first run q2 adopting the Solar parameters as a first guess for each star.

\item After q2 has converged to a set of stellar parameters, the differential abundances relative to the Sun are calculated for the following elements: C I, Na I, Mg I, Al I, Si I, S I, Ca I, Sc II, Ti I, Ti II, V I, Cr I, Cr II, Mn I, Fe I, Fe II, Co I, Ni I, Cu I, Zn I, Y II, Zr II, and Ba II. Through the blends driver in MOOG and adopting the line list from the Kurucz database, the q2 code corrects the abundances of V, Mn, Co, Cu, and Y for hyperfine splitting effects, by using the HFS components in the input line list. For each element, we perform a 3-sigma clipping on the abundances yielded by each EW measurement. This allows us to remove the EW measurements affected by telluric lines or unresolved blendings.

\item A second run of q2 is performed using the cleaned list of EW measurements. This second iteration yields the final atmospheric parameters listed in Table~\ref{HARPS_param}, which also reports the parallaxes and proper motions from Gaia eDR3 \cite{Lindegren20}. With these final atmospheric parameters, we repeat the calculation of the differential abundances relative to the Sun. The resulting abundances are listed in Table~\ref{HARPS_abu_sun}, together with their uncertainties and the number of lines used for the abundance determinations. The error budget associated with each elemental abundance is obtained by summing in quadrature the standard error of the mean among the lines, and the propagated effects of the uncertainties on the stellar parameters. We also determine the differential abundances within each pair, which are listed in Table~\ref{HARPS_diff_abu}.

\item Finally, we employ the stellar parameters and abundances listed in Tables~\ref{HARPS_param} and \ref{HARPS_abu_sun} to measure the differential Li abundances used in Figure~\ref{Lithium}. This analysis is performed through the spectral synthesis of the Li line at 6707.75~$\AA$, following a procedure already tested in our previous works \cite{Carlos19}. Through the same technique we also measure differential Li abundances for the stellar sample of Nagar et al. (2020) \cite{Nagar20} and in the binary pair HAT-P-1A/B . All abundances employed in Figure~\ref{Lithium}, including those measured in this work, are listed in Table~\ref{Tab:lithium}.

\end{itemize}

{\centering
\subsection*{C. Probability of finding anomalous pairs}
}

The probability P$_{\rm Anom}$ is modelled with a Sigmoid function of $\langle$T$_{\rm eff}$$\rangle$, defined as follows:

\begin{equation}
\label{sigmoid}
   P_{\rm Anom} = \frac{\alpha}{1+e^{-\frac{\beta (\langle T_{\rm eff}\rangle - k)}{10^3 K}}}.
\end{equation}

The k parameter indicates the $\langle$T$_{\rm eff}$$\rangle$ at which the knee of the Sigmoid function is located, $\alpha$ defines the highest probability value, and $\beta$ defines the sign and the steepness of the possible correlation between P$_{\rm Anom}$ and $\langle$T$_{\rm eff}$$\rangle$. Namely, when $\beta$ is positive there is a positive correlation between P$_{\rm Anom}$ and $\langle$T$_{\rm eff}$$\rangle$, while a $\beta$ equal to zero indicates that the two quantities are not correlated.

We sample the probability distribution of these parameters through a Markov Chain Monte Carlo (MCMC) simulation. The Bayesian inference is conditioned on the sample of 107 binary systems through a Bernulli random variable with probability equal to P$_{\rm Anom}$. Normal probability distributions $\mathcal{N}$($\mu$,$\sigma$) are chosen as the priors of the three parameters k, $\alpha$, and $\beta$. Namely, the k prior is equal to $\mathcal{N}$(T$_{\rm eff \odot}$,50), while the $\alpha$ prior is $\mathcal{N}$($f_{Anom}$,0.2) where $f_{Anom}$ is defined as the ratio between the number of chemically anomalous pairs (33) and of the total sample (107) and is equal to 0.308. The $\alpha$ prior is also bounded between 0 and 1. Finally, the $\beta$ prior is $\mathcal{N}$(0,10). Note that this latter is centred at zero, therefore the MCMC simulation is initially ``agnostic'' on the type of correlation between P$_{\rm Anom}$ and $\langle$T$_{\rm eff}$$\rangle$. We run the simulation with 20,000 walkers, using the Metropolis sampler algorithm, and a burn-in of 10,000. The script is written in Python using the $\tt{pymc3}$ package \cite{Salvatier16}. The correct convergence of the simulation is checked against the traces and the autocorrelation function plots. The 5-, 50-, and 95-percentiles of the posterior distributions are summarised in Table~\ref{Tab:posteriors}. We note that, while the $\beta$ prior was centred on zero, the simulation selects positive $\beta$ values and converges to a set of $\beta$ solutions that are for the 95$\%$ larger than 1.96. That strongly indicates the presence of a positive relation between P$_{\rm Anom}$ and $\langle$T$_{\rm eff}$$\rangle$. These posteriors are used to calculate the 90$\%$ confidence interval of P$_{\rm Anom}$, which is shown in Figure~\ref{prob_estimate}.

\begin{figure}[h]
\begin{center}
 \includegraphics[width=\textwidth]{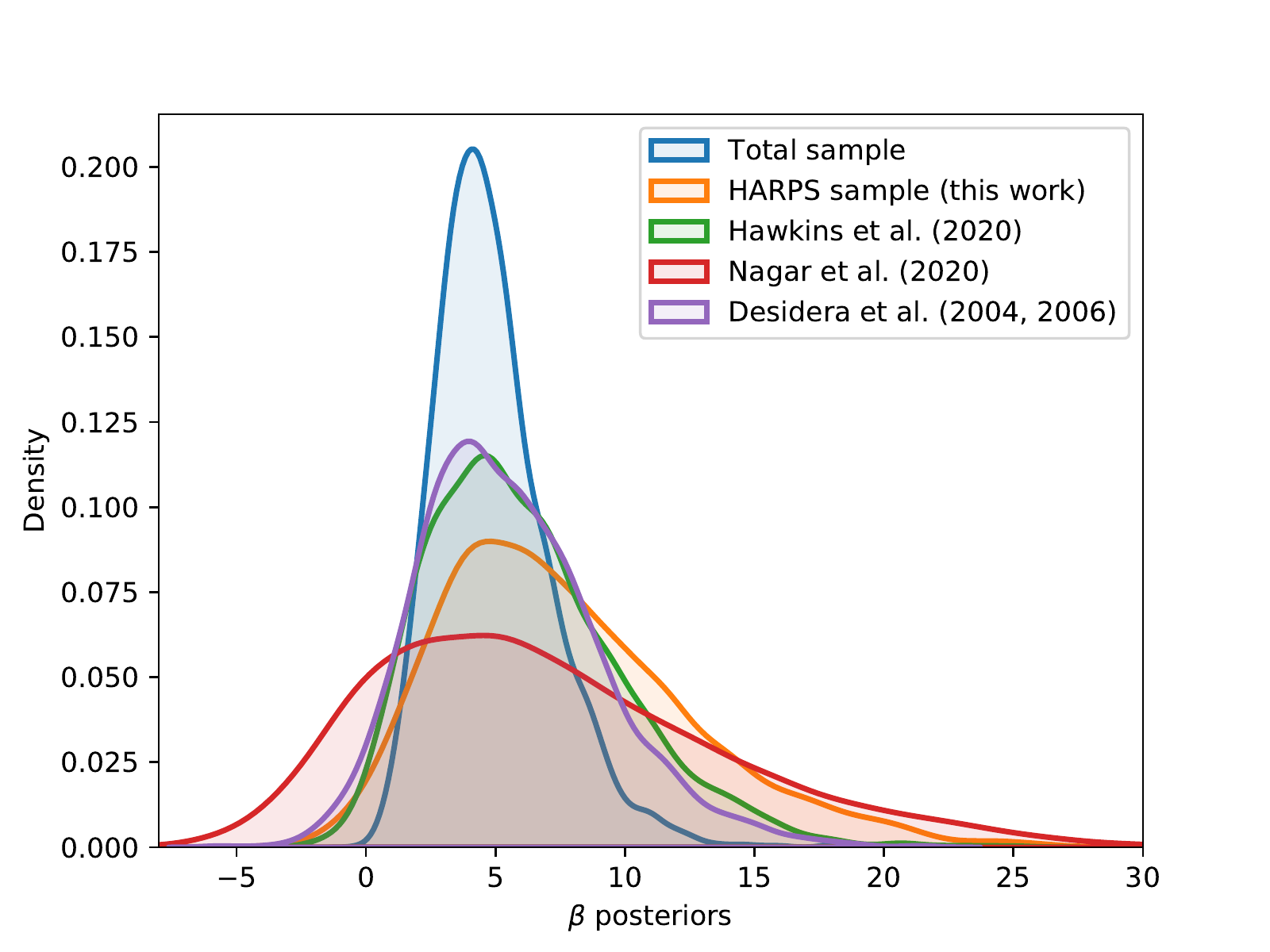}
  \end{center}
 \caption{\textbf{Consistency test of individual subsamples.} The figure shows the $\beta$ posterior distributions obtained from MCMC simulations conditioned individually on four subsamples. The posterior distributions obtained from the entire sample of 107 binary pairs is also shown (blue curve). These is a general agreement among the allo posterior distributions, which indicates that the individual subsamples are consistently leading to similar results.}
 \label{posteriors_checks}
\end{figure}

Since in this analysis we combine abundances from multiple sources, we test how the different datasets are affecting our results. In particular, we want to verify whether all the individual samples are consistently leading to similar conclusions or instead one or two samples alone are conditioning our results against all the others. To do so, we repeat the analysis described above for four individual samples: i) the HARPS sample analysed in this work; ii) the sample from Hawkins et al. (2020) \cite{Hawkins20}; iii) the sample from Nagar et al. (2020) \cite{Nagar20}; iv) the sample from Desidera et al. (2004, 2006) \cite{Desidera04,Desidera06}. Fig.~\ref{posteriors_checks} shows the resulting $\beta$ posterior distributions resulting from these four individual samples together with the $\beta$ posterior distribution obtained from the total sample. Here we observe a general agreement among the posterior distributions and, considering that the $\beta$ prior was centred on zero, we conclude that the individual samples consistently favour positive $\beta$ values.

Figure~\ref{prob_estimate} also shows with dashed lines the probabilities derived from three mock datasets. Each of these latter are composed by 500 pairs of identical stars with age equal to 1 Gyr and masses that can range between 0.66 and 1.4 M$_{\odot}$. All these stars are in their main-sequence phase and have a stable convective zone. Then, after we randomly pollute the convective zone (CZ) of the 27$\%$ of these stars with planetary material, we calculate the differential iron abundances between the pairs $\Delta$Fe. For this calculation we assume the mass of the CZ from the Yonsei-Yale stellar models \cite{Spada17}, a stellar abundance of iron relative to other metals as in Asplund et al. (2006) \cite{Asplund06} and the planet iron abundance from Wang et al. (2018) \cite{Wang18}. The amount of planetary material that is injected into the stellar CZ is randomly drawn from a Gaussian distribution centred at  1.5 M$_{\Earth}$, with a standard deviation of 1.2 M$_{\Earth}$ and ranging between 0.1 and 20 M$_{\Earth}$. The metallicity of the planet scales with the stellar metallicity. The three mock datasets are built separately from models with metallicities of 0.016 (solar metallicity), 0.005 (metal poor), and 0.032 (metal rich). The 500 pairs of each mock dataset are then classified as \textit{chemically anomalous} when their $\Delta$Fe$>$0.02 dex, otherwise they are flagged as \textit{chemically homogeneous}. Finally, the three probability functions shown in Figure~\ref{prob_estimate} are obtained through a MCMC simulation as described above.

{\centering
\subsection*{D. [Fe/C] percentile ranks}
}
The [Fe/C] percentile ranks allow us to compare the refractory-to-volatile ratios of the stars in the \textit{chemically anomalous} pairs to the typical values of coeval solar twins. For this analysis we initially considered 17 \textit{chemically anomalous} pairs (34 stars) belonging to the first group (see Section~A) and for which both [C/H] and [Fe/H] are known. The ages of these stars are listed in Table~\ref{Tab:ages}. They are calculated through the same method, grid of isochrones (i.e., Yonsei-Yale; \cite{Spada17}) and tool (i.e., q2; \cite{Ramirez14}) employed in Casali et al. (2020) \cite{Casali20}. The comparison sample of solar twins was also taken from Casali et al. (2020) \cite{Casali20}.

For each component of the binary systems we selected an \textit{ad hoc} comparison sample of solar twins in order to limit the impact of Galactic Chemical Evolution in our analysis. The comparison sample is exclusively composed by solar twins having ages within 2 Gyr of the age of the binary component. Similarly, we imposed that the comparison sample is composed by solar twins having [Fe/H] within 0.1 dex of the [Fe/H] of the binary component. This allows us to compare stars that are nearly coeval and that formed at similar Galactic evolution. Out of the initial 34 stars we discarded 3 stars (i.e., XO-2N/S and SW1042-0350B) that have comparison samples smaller than 5 solar twins.

The comparison is performed as follows. For each of the 31 stars we randomly draw 1,000 values of [Fe/H] and [C/H] abundances from the normal distributions $\mathcal{N}$([X/H],$\sigma_{[X/H]}$) where X is either Fe or C. Then, from these values, we obtain 1,000 [Fe/C] abundance ratios that we use to calculate as many percentile ranks through the comparison with the [Fe/C] distribution of the solar twins selected as above. The full distribution of the resulting 31,000 [Fe/C] percentile scores calculated for the 31 stars is shown in Figure~\ref{volatiles_refractories}, separately for the metal-rich and metal-poor components of the binary pairs.

\subsection*{Data availability}
All data generated or analysed during this study are included in this published article as Source Data files.

\newpage
\bibliographystyle{naturemag}
\bibliography{Bibliography.bib}

\newpage
\subsection*{Acknowledgements}
This work has made use of observations collected at the European Southern Observatory (ESO programmes 188.C-0265, 0103.C-0785, and 0101.C-0275) and of data from the European Space Agency (ESA) mission Gaia. We are grateful to K. Hawkins, and F. Liu for having shared with us tabular and spectroscopic data. LS thanks A.I. Karakas for her support during the project and acknowledges financial support from the Australian Research
Council (discovery Project 170100521) and from the Australian Research Council Centre of Excellence for All Sky Astrophysics in 3 Dimensions (ASTRO 3D), through project number CE170100013. JM thanks FAPESP (2018/04055-8).

\subsection*{Author Contributions}

All authors assisted in the development, analysis and writing of the paper. LS led and played a part in all aspects of the analysis, and wrote the manuscript. PS carried out the spectroscopic analysis of the HARPS sample. JM and MB contributed in designing this study and in the interpretations of the results. ARC identified the binary pairs with HARPS spectra in the ESO archive and provided the cross-match to the Gaia dataset. MC and EF derived the Li abundances. AV contributed to the discussion of the results.

\subsection*{Correspondence}
Correspondence and requests for materials should be addressed to LS (spina.astro \textit{at} gmail.com).

\subsection*{Declaration of Interests}
The authors declare no competing interests.

\newpage
\section*{Tables}

\begin{table}[h]
\caption{Binary paris. - Full table available online at the CDS.}
\centering
\label{lit_data}
{\small
\begin{tabular}{ccccc}
\hline
A component & B component & $\langle$T$_{\rm eff}$$\rangle$ & $\Delta$Fe & Ref.  \\ \hline
1448493530351691520 & 1448493427272476288 & 6410.0 & -0.100$\pm$0.014 & \cite{Hawkins20}\\
219605599154126976 & 219593745044391552 & 6450.5 & 0.010$\pm$0.022 & \cite{Hawkins20}\\
232899966044906496 & 232899966044905472 & 5857.0 & 0.040$\pm$0.014 & \cite{Hawkins20}\\
238164255921243776 & 238163534366737792 & 6011.0 & -0.110$\pm$0.014 & \cite{Hawkins20}\\
2493516351151864960 & 2493516351151865088 & 6276.5 & 0.010$\pm$0.014 & \cite{Hawkins20}\\
... & ... & ... & ... & ...\\ \hline
\end{tabular}
}
\end{table}

\begin{table}[h]
\caption{HARPS sample. Atmospheric and astrometric parameters. - Full table available online at the CDS.}
\centering
\label{HARPS_param}
{\tiny
\begin{tabular}{cccccccccc}
\hline
Star & R.A. & Dec. & T$_{\rm eff}$ & logg & [Fe/H] & v$_{\rm t}$ & Plx & pmRA & pmDec \\
 & [deg] & [deg] & [K] & [dex] & [dex] & [km s$^{-1}$] & [mas] & [mas yr$^{-1}$] & [mas yr$^{-1}$] \\ \hline
CD-2616866 & 0.0404928 & -25.3248377 & 6237$\pm$54 & 4.070$\pm$0.114 & -0.089$\pm$0.034 & 1.90$\pm$0.10 & 3.421$\pm$0.016 & 0.792$\pm$0.018 & -8.395$\pm$0.013 \\
CD-2616866B & 0.0409676 & -25.3225333 & 6200$\pm$35 & 3.888$\pm$0.111 & -0.112$\pm$0.025 & 1.82$\pm$0.07 & 3.393$\pm$0.018 & 1.067$\pm$0.022 & -11.916$\pm$0.014 \\
HD4552 & 11.9218787 & 12.8791284 & 6278$\pm$23 & 4.323$\pm$0.056 & -0.005$\pm$0.016 & 1.64$\pm$0.03 & 9.188$\pm$0.016 & 84.485$\pm$0.021 & -35.236$\pm$0.016 \\
BD+120090 & 11.9322262 & 12.8720021 & 5994$\pm$6 & 4.298$\pm$0.016 & -0.011$\pm$0.005 & 1.28$\pm$0.01 & 9.229$\pm$0.017 & 84.098$\pm$0.020 & -35.772$\pm$0.015 \\
CD-50524 & 28.1066953 & -49.5344249 & 6224$\pm$9 & 4.385$\pm$0.029 & -0.091$\pm$0.007 & 1.52$\pm$0.02 & 8.598$\pm$0.013 & 32.434$\pm$0.012 & -48.125$\pm$0.013 \\
... & ... & ... & ... & ... & ... & ... & ... & ... & ... \\ \hline
\end{tabular}
}
\end{table}

\begin{table}[h]
\caption{HARPS sample. Chemical abundances relative to the Sun. - Full table available online at the CDS.}
\centering
\label{HARPS_abu_sun}
{\tiny
\begin{tabular}{cccccccc}
\hline
Star & [CI/H] & [NaI/H] & [MgI/H] & [AlI/H] & [SiI/H] & [SI/H] & ... \\ \hline
CD-2616866 & -0.076$\pm$0.059 (3) & -0.196$\pm$0.049 (2) & -0.069$\pm$0.040 (3) & 0.126$\pm$0.019 (1) & -0.012$\pm$0.033 (14) & -0.172$\pm$0.043 (1) & ... \\
CD-2616866B & 0.020$\pm$0.053 (3) & -0.045$\pm$0.042 (2 )& -0.067$\pm$0.027 (3) & -0.198$\pm$0.013 (1) & -0.026$\pm$0.019 (11) & -0.138$\pm$0.049 (3) & ...\\
HD4552 & 0.024$\pm$0.032 (2) & -0.005$\pm$0.012 (2) & 0.013$\pm$0.013 (2) & -0.024$\pm$0.009 (1) & 0.036$\pm$0.012 (11) & -0.082$\pm$0.044 (3) & ... \\
BD+120090 & -0.012$\pm$0.008 (3) & 0.019$\pm$0.014 (3) & 0.001$\pm$0.009 (5) & 0.005$\pm$0.003 (2) & 0.009$\pm$0.003 (14) & -0.024$\pm$0.030 (4) & ...\\
CD-50524 & -0.088$\pm$0.016 (3) & -0.075$\pm$0.045 (3) & -0.050$\pm$0.020 (4) & -0.150$\pm$0.011 (2) & -0.059$\pm$0.006 (14) & -0.101$\pm$0.046 (4) & ...\\
... & ... & ... & ... & ... & ... & ... & ...  \\ \hline
\end{tabular}
}
\end{table}

\begin{table}[h]
\caption{HARPS sample. Differential abundances within the pair. - Full table available online at the CDS.}
\centering
\label{HARPS_diff_abu}
{\tiny
\begin{tabular}{ccccccccc}
\hline
Component A & Component B & $\Delta$CI & $\Delta$NaI & $\Delta$MgI & $\Delta$AlI & $\Delta$SiI & $\Delta$SI & ... \\ \hline
BD-104948A & BD-104948B & 0.000$\pm$0.017 (2) & -0.008$\pm$0.020 (3) & -0.017$\pm$0.011 (5) & 0.001$\pm$0.022 (2) & -0.006$\pm$0.005 (14) & -0.058$\pm$0.034 (4) & ...\\
BD+10303B & HD13904 & -0.027$\pm$0.018 (1) & 0.011$\pm$0.018 (2) & -0.010$\pm$0.029 (4) & -0.116$\pm$0.052 (2) & 0.017$\pm$0.020 (14) & -0.115$\pm$0.018 (2) & ...\\
BD+120090 & HD4552 & 0.034,0.029 (2) & -0.010,0.012 (2) & 0.003,0.016 (2) & -0.027,0.009 (1) & 0.025,0.012 (11) & -0.034,0.030 (3) & ...\\
CD-2616866 & CD-2616866B & 0.096$\pm$0.072 (3) & 0.151$\pm$0.025 (2) & 0.001$\pm$0.060 (3) & -0.324$\pm$0.023 (1) & -0.004$\pm$0.026 (11) & 0.084$\pm$0.051 (1) & ...\\
CD-3112467B & HD143336 & 0.014$\pm$0.027 (2) & 0.087$\pm$0.009 (1) & 0.086$\pm$0.026 (5) & --- & 0.048$\pm$0.015 (14) & -0.026$\pm$0.061 (2) & ...\\
... & ... & ... & ... & ... & ... & ... & ... & ...  \\ \hline
\end{tabular}
}
\end{table}

\begin{table*}
\caption{Differential Li abundances.}
\centering
\label{Tab:lithium}
{\small
\begin{tabular}{cccc}
\hline
Component A & Component B   & $\Delta$Li &Ref. \\
 & & [dex] &  \\ \hline
16 Cygni A  & 16 Cygni B  & -0.70$\pm$0.04 & \cite{TucciMaia19} \\
Kronos  & Krios  & -0.50$\pm$0.07 & \cite{Oh18} \\
HAT-P-1A  & HAT-P-1B  & -0.30$\pm$0.05 & This work \\
HAT-P-4A  & HAT-P-4B  & -0.30$\pm$0.05 & \cite{Saffe17} \\
HIP39409A  & HIP39409B  & -0.0$\pm$0.3 & This work \\
HR4443  & HR4444  & -0.27$\pm$0.13 & This work \\
HIP44858 & HIP44864  & 0.00$\pm$0.03 & This work \\
HIP58298A & HIP58298B  & 0.07$\pm$0.04 & This work \\
HIP47836 & HIP47839  & 0.09$\pm$0.03 & This work \\
HD98744 & HD98745  & 0.91$\pm$0.12 & This work \\
HIP70269A & HIP70269B  & 0.00$\pm$0.03 & This work \\
HD105421 & HD105422  & -1.43$\pm$0.12 & This work \\
HIP70386A & HIP70386B  & -0.16$\pm$0.02 & This work \\
HD111484A & HD111484B  & 0.18$\pm$0.03 & This work \\
HD189739  & HD189760  & -0.08$\pm$0.04 & This work \\
CD-4714611B  & HD221550  & 0.01$\pm$0.13 & This work \\
HD206429  & HD206428  & 1.9$\pm$0.4 & This work \\
HD20439  & HD20430  & 0.22$\pm$0.02 & This work \\
HIP49520A  & HIP49520B  & -0.12$\pm$0.04 & This work \\
HD29167  & CPD-60315B  & 0.00$\pm$0.09 & This work \\
HD196068  & HD196067  & -0.16$\pm$0.03 & This work \\
HD96273  & BD+07-2411B  & 0.63$\pm$0.4 & This work \\
HD147722  & HD147723  & -0.19$\pm$0.03 & This work \\
HD181544  & HD181428  & 0.01$\pm$0.04 & This work \\
HIP34407  & HD54100  & -0.02$\pm$0.04 & This work \\
CD-3112467B  & HD143336  & 0.33$\pm$0.11 & This work \\
HIP104687A  & HIP104687B  & -0.04$\pm$0.03 & This work \\
BD-104948A  & BD-104948B  & 0.22$\pm$0.06 & This work \\
HD117190  & HD117190B  & -0.24$\pm$0.08 & This work \\
HD133131A  & HD133131B  & 0.1$\pm$0.3 & This work \\
HD182817  & HD182797  & 0.06$\pm$0.04 & This work \\
CD-50524  & HD11584  & 0.09$\pm$0.04 & This work \\
HD24085  & HD24062  & 0.13$\pm$0.02 & This work \\ \hline
\end{tabular}
}
\end{table*}

\begin{table*}
\caption{Percentiles of the posterior distributions.}
\centering
\label{Tab:posteriors}
{\small
\begin{tabular}{cccc}
\hline
Parameters & 5$\%$ & 50$\%$ & 95$\%$ \\ \hline
k [K] & 5677 & 5757 & 5840\\
$\alpha$ & 0.38 & 0.50 & 0.62 \\
$\beta$ & 1.96 & 4.63 & 8.93 \\ \hline
\end{tabular}
}
\end{table*}

\begin{table*}
\caption{Stellar ages.}
\centering
\label{Tab:ages}
{\small
\begin{tabular}{ccccc}
\hline
Star & Age \\ 
 & [Gyr] \\ \hline
XO-2N & 13.9$^{+0.6}_{-1.8}$ \\  
XO-2S & 11.3$^{+1.0}_{-1.6}$ \\  
16 Cygni A & 6.6$^{+0.5}_{-0.4}$ \\  
16 Cygni B & 6.8$^{+0.4}_{-0.6}$ \\  
Kronos & 2.7$^{+1.1}_{-1.2}$ \\  
Krios & 6.7$^{+0.7}_{-0.9}$ \\  
HAT-P-4A & 3.4$^{+0.7}_{-1.8}$ \\  
HAT-P-4B & 3.4$^{+0.6}_{-1.9}$ \\  
HD98744 & 5.2$^{+0.5}_{-0.7}$ \\  
HD98745 & 4.9$^{+0.6}_{-1.7}$ \\  
HD105421 & 0.8$^{+0.8}_{-0.4}$ \\  
HD105422 & 1.2$^{+1.3}_{-0.6}$ \\  
SW1042-0350B & 14.4$^{+0.4}_{-1.0}$ \\  
SW1042-0350 & 9.8$^{+0.4}_{-0.4}$ \\  
HD20430 & 0.7$^{+0.2}_{-0.3}$ \\  
HD20439 & 0.7$^{+0.3}_{-0.3}$ \\  
HD196068 & 3.7$^{+0.3}_{-0.4}$ \\  
HD196067 & 3.9$^{+0.4}_{-0.3}$ \\  
HD96273 & 4.0$^{+0.4}_{-0.5}$ \\  
BD+07-2411B & 3.9$^{+1.2}_{-1.5}$ \\  
HD181544 & 4.6$^{+0.4}_{-0.4}$ \\  
HD181428 & 3.9$^{+0.4}_{-0.3}$ \\  
HD169392A & 5.4$^{+0.4}_{-0.4}$ \\  
HD169392B & 5.7$^{+0.7}_{-0.7}$ \\  
HIP34407 & 7.6$^{+0.7}_{-1.1}$ \\  
HD54100 & 8.0$^{+0.6}_{-0.7}$ \\  
CD-3112467B & 2.1$^{+1.5}_{-1.1}$ \\  
HD143336 & 3.1$^{+0.9}_{-1.5}$ \\  
CD-50524 & 1.8$^{+0.7}_{-0.7}$ \\  
HD11584 & 2.1$^{+0.6}_{-0.6}$ \\  
HD24085 & 3.4$^{+0.4}_{-0.3}$ \\  
HD24062 & 3.9$^{+0.5}_{-0.2}$ \\  
HD192343 & 7.3$^{+0.3}_{-0.4}$ \\  
HD192344 & 8.0$^{+0.3}_{-0.5}$ \\  \hline 
\end{tabular}
}
\end{table*}

\end{document}